\documentstyle[epsf,epsfig,amstex]{europhys}

\newif\ifboo \boofalse

\begin{document}

\euro{57}{5}{625-631}{2002}
\Date{1 March 2002}
\shorttitle{C.-C.~Lo {\it et al.}: DYNAMICS OF SLEEP-WAKE 
TRANSITIONS}
\title{
Dynamics of Sleep-Wake Transitions During Sleep\\
}

\author{C.-C.~Lo\inst{1},
L.~A.~Nunes~Amaral\inst{1,2}, S.~Havlin\inst{3}, P.~Ch.~Ivanov\inst{1,2},\\
T.~Penzel\inst{4}, J.-H.~Peter\inst{4}, and H. E.~Stanley\inst{1}}
\institute{\inst{1} Center for Polymer Studies and Department of Physics,
       Boston University, \\ Boston, MA 02215, USA \\
     \inst{2} Cardiovascular Division, Beth Israel Deaconess Medical
       Center, \\
       Harvard Medical School, Boston, MA 02215, USA \\
     \inst{3} Gonda Goldschmid Center and Department of Physics, 
    Bar-Ilan University, \\Ramat Gan, Israel\\
     \inst{4} Klinik f\"ur Innere Medizin, Philipps-Universit\"at, 
      \\Baldingerstrasse 1, Marburg D-35033, Germany \\ }
\rec{11 September 2001}{in the final form 29 November 2001}
\pacs{
\Pacs{02}{50.Ey}{Stochastic processes}
\Pacs{05}{40.Fb}{Random walks and Levy flights}
\Pacs{05}{40.-a}{Fluctuation phenomena, random processes, noise, and Brownian motion}
}

\maketitle


\begin{abstract}
We study the dynamics of the awakening during the night
for healthy subjects and find that the wake and the sleep 
periods exhibit completely different behavior:
the durations of wake periods are characterized by a scale-free 
power-law distribution, while the durations of sleep periods
have an exponential distribution with
a characteristic time scale. We find that the characteristic 
time scale of sleep periods changes throughout the night.
In contrast, there is no measurable 
variation in the power-law behavior for the durations of wake periods.
We develop a stochastic model
which agrees with the data and suggests that the difference in the 
dynamics of sleep and wake states arises from the constraints on the
number of microstates in the sleep-wake system.
\end{abstract}

\begin{figure}
\begin{center}
 \begin{minipage}[b]{0.95\linewidth}
 \epsfig{file=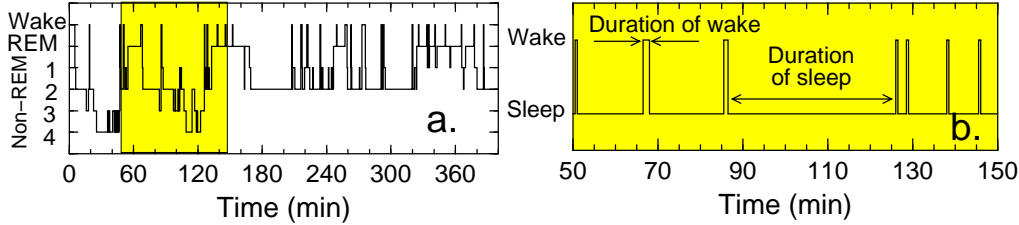, width=\linewidth}
\end{minipage}
\end{center}
\vspace{-0.4cm}
\caption{
Sleep-wake transitions during nocturnal sleep.  (a) Representative example of
sleep-stage transitions from a healthy subject.  Data were recorded in a
sleep laboratory according to the Rechtschaffen and Kales criteria
\protect\cite{Rechtschaffen-A-1968a}: two channels of electroencephalography
(EEG), two channels of electrooculography (EOG) and one channel of submental
electromyography (EMG) were recorded. Signals were digitized at 100 Hz and 12
bit resolution, and visually ``scored'' by sleep experts in segments of 30
seconds for sleep stages: wakefulness, rapid-eye-movement (REM) sleep and
non-REM sleep stages 1, 2, 3 and 4.  (b) Magnification of the shaded region
in (a). In order to study sleep-wake transitions, we reduce the five sleep 
stages into a single sleep state by grouping rapid-eye-movement (REM) sleep 
and sleep stages 1 to 4 into a single sleep state.}
\vspace{-0.2cm}
\label{f.sleep_stages}
\end{figure}

In clinical sleep centers, the ``total sleep time'' and the
``total wake time'' during the night are used to evaluate
sleep efficacy and to diagnose sleep disorders. 
However, the total wake time during a long period of 
nocturnal sleep is actually comprised of
many short wake intervals (Fig.~\ref{f.sleep_stages}).
This fact suggests that the ``total wake time'' during sleep 
is not sufficient to characterize the complex sleep-wake transitions
and that it is important to ask how 
periods of the wake state distribute during the course of the night.
Although recent studies have focused on sleep control at the 
neuronal level 
\cite{Chicurel-M-2000a,McGinty-D-2000a,Benington-J-2000a,Gallopin-T-2000a},
very little is known about the dynamical mechanisms responsible for the time
structure or even the statistics of the abrupt sleep-wake transitions
during the night. Furthermore, different scaling
behavior between sleep and wake activity and between different sleep 
stages has been observed \cite{Ivanov-P-1999b,Bunde-A-2000a}.
Hence, investigating the statistical properties of the wake 
and sleep states throughout the night may provide not only a more 
informative measure but also insight into the 
mechanisms of the sleep-wake transition.


We analyze 39 full-night sleep records collected from 20 healthy 
subjects (11 females and 9 males, ages 23--57, with average sleep 
duration 7.0 hours). We first study the distribution of durations 
of the sleep and of the wake states during the night (Fig.~\ref{f.sleep_stages}).  
We calculate the cumulative distribution of durations, defined as
\begin{equation}
P(t) \equiv \int_t^\infty p(r) dr \,,
\label{eq.cumulative}
\end{equation}
where $p(r)$ is the probability density function of durations between
$r$ and $r+dr$.
We analyze $P(t)$ of the wake state, and we find that the data follow a 
power-law distribution, 
\begin{equation}
P(t)\sim t^{-\alpha} \,.
\label{eq.power-law} 
\end{equation}
We calculate the exponent $\alpha$ for each of the 20 subjects, and find 
an average exponent $\alpha = 1.3$ with a standard deviation $\sigma=0.4$. 

\begin{figure}
\begin{center}
\begin{minipage}[b]{0.35\linewidth}
\epsfig{file=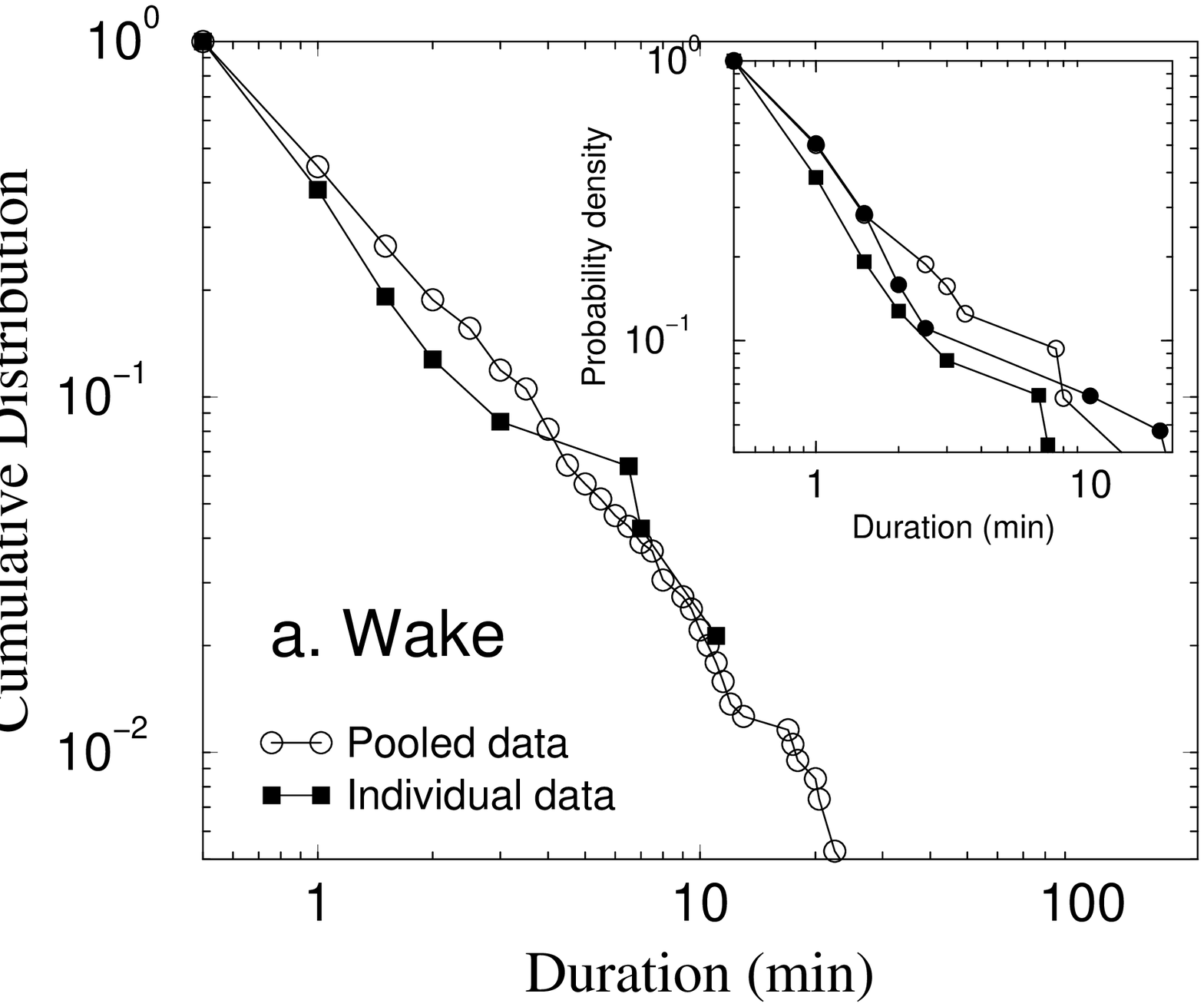, width=\linewidth}
\end{minipage}\hspace{0.5cm}
\begin{minipage}[b]{0.35\linewidth}
\epsfig{file=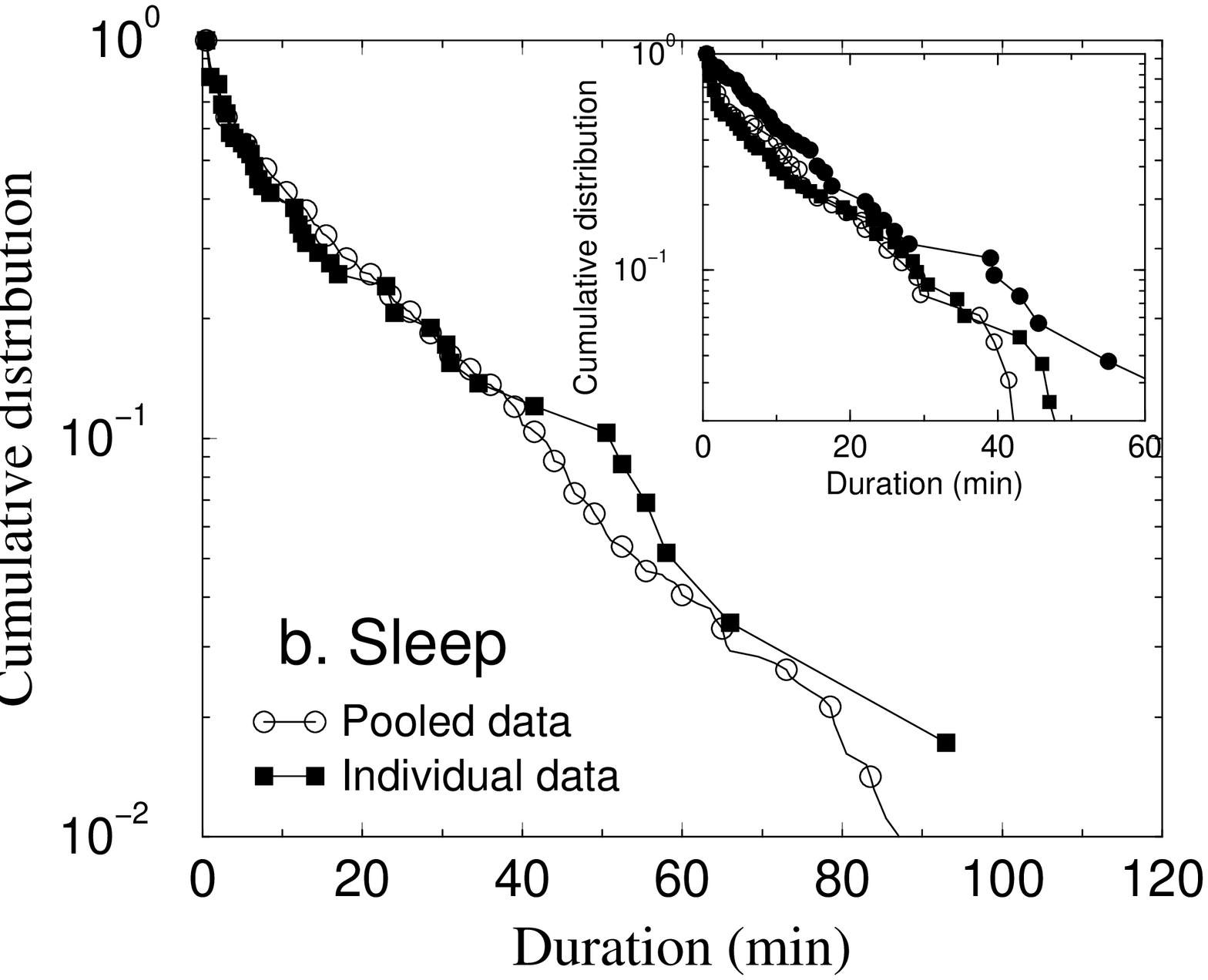, width=\linewidth}
\end{minipage}
\end{center}
\vspace{-0.4cm}
\caption{
Cumulative probability distribution $P(t)$ of sleep and wake durations
of individual and pooled data. Double-logarithmic plot of $P(t)$ 
of wake durations (a) and semi-logarithmic plot of $P(t)$ of sleep 
durations (b) for pooled data and for data from one typical 
subject. $P(t)$ for three typical subjects is shown in the insets. 
Note that due to 
limited number of sleep-wake periods for each subject, it is difficult 
to determine the functional form for individual subjects. We perform 
K-S test and compare the probability density $p(t)$ for all  
individual data sets and pooled data for both wake and sleep periods.
For both sleep and wake, less than $10\%$ of the individual data
fall below the $0.05$ significant level of disproof of the null
hypothesis, that $p(t)$ for each individual subject is very 
likely drawn from the same distribution. The K-S statistics significantly
improves if we use recordings only from the second night.
Therefore, pooling all data improves the
statistics by preserving the form of $p(t)$.
}
\label{f.pdf}
\vspace{-0.2cm}
\end{figure}

\begin{figure}
\begin{center}
\begin{minipage}[b]{0.35\linewidth}
\epsfig{file=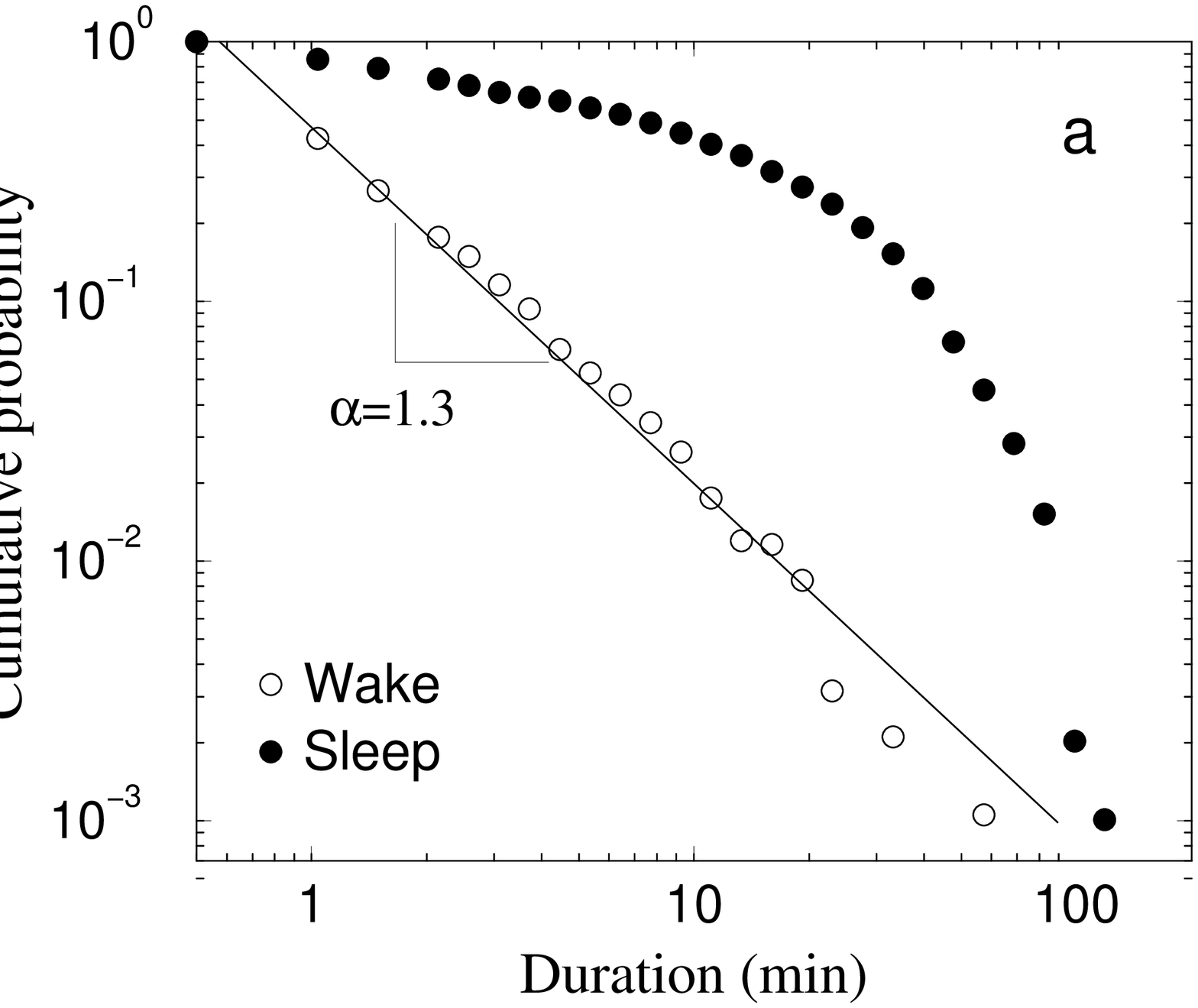, width=\linewidth}
\end{minipage}\hspace{0.5cm}
\begin{minipage}[b]{0.35\linewidth}
\epsfig{file=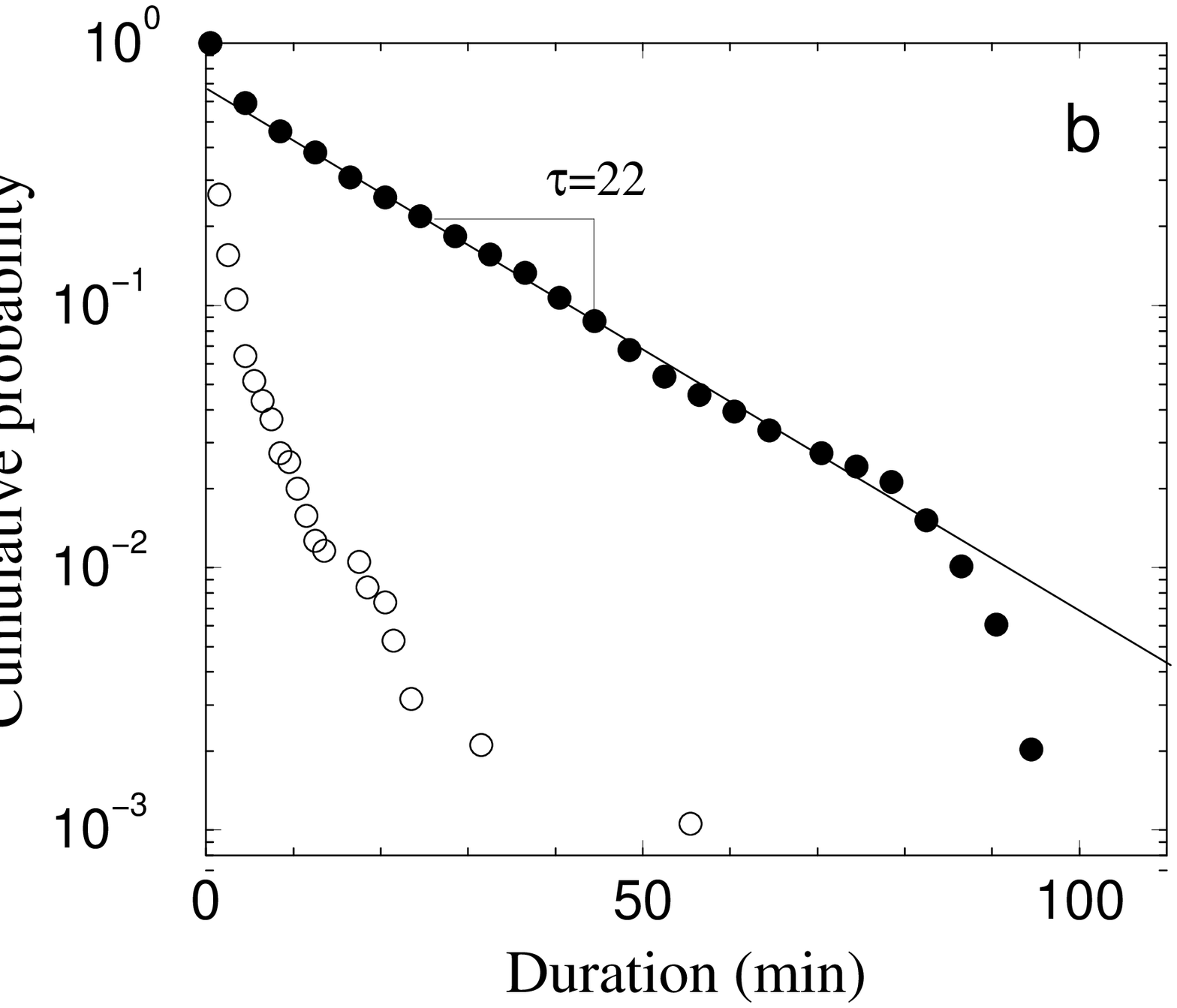, width=\linewidth}
\end{minipage}
\end{center}
\vspace{-0.4cm}
\caption{ Cumulative distribution of durations $P(t)$ of sleep and wake 
states from data.  (a) Double-logarithmic plot of $P(t)$ from
the pooled data. For the wake state, the distribution closely follows a 
straight line with a slope $\alpha=1.3 \pm 0.1$, indicating
power-law behavior of the form of Eq.~(\protect\ref{eq.power-law}).
(b) Semi-logarithmic plot of $P(t)$.  For
the sleep state, the distribution follows a straight line with a slope $1/\tau$
where $\tau=22 \pm1 $, indicating an
exponential behavior of the form of Eq.~(\protect\ref{eq.exponential}).
It has been reported that the individual sleep stages have exponential
distributions of durations
\protect\cite{Williams-R-1964a,Brezinova-V-1975a,Kemp-B-1986a}. 
Hence we expect an
exponential distribution of durations for the sleep state.
}
\vspace{-0.2cm}
\label{f.data}
\end{figure}

It is important to verify that the data 
from individual records correspond to the same probability distribution.
To this 
end, we apply the Kolmogorov-Smirnov test to the data from individual records.
We find that we cannot reject the null hypothesis that $p(t)$ of the
wake state of each subject is drawn from the same distribution, 
suggesting that one can pool all data together to improve statistics without 
changing the distribution (Fig. ~\ref{f.pdf}a). Pooling the data 
from all 39 records, we find that 
$P(t)$ of the wake state is consistent with a  power-law distribution
with an exponent $\alpha = 1.3 \pm 0.1$ (Fig.~\ref{f.data}a).

In order to verify that the distribution of durations of wake state is 
better described by a power law 
rather than an exponential or a stretched exponential 
functional forms, we fit these curves to the distributions from pooled 
data. Using Levenberg-Marquardt method, we find that both exponential 
and stretched exponential forms lead to worse fits. 
The $\chi^2$ errors of power-law fit, exponential fit
and stretched exponential fit are $3 \times 10^{-5}$, $1.6 \times 10^{-3}$ 
and $3.5 \times 10^{-3}$, respectively. We also check the results by plotting
(i) $\log P(t)$ versus $t$ and
(ii) $\log(|\log P(t)|)$ versus $\log t$
\footnote{For the stretched exponential $y=a\exp{(-bx^c)}$, where a, b and c
are constants, the $\log(|\log y|)$ versus $\log x$ plot is not a straight
line unless $a=1$. Since we don't know what the corresponding
value of $a$ is in our data, we can not rescale $y$ so that $a=1$.
The solution is to shift $x$ for a certain value to make $y=1$ when $x=0$,
in which case $a=1$. In our data, $P(t)=1$ when $t=0.5$, so we shift $t$ by
$-0.5$ before plotting $\log(|\log P(t)|)$ versus $\log t$.}
and find in both cases that the data are 
clearly more curved than when we plot $\log P(t)$ versus $\log t$,
indicating that a power law provides the best description of the data
\footnote{According Eq.~\ref{eq.cumulative}, if $P(t)$ is a power-law
function, so is $p(t)$. We also separately check the functional form 
of $p(t)$ for
the data with same procedure and find that the power law provides the 
best description of the data.}.

We perform a similar analysis for the sleep state and find, in contrast 
to the result for the wake state, that the data in large time region
($t > 5$ min) exhibit exponential behavior 
\begin{equation}
P(t)\sim e^{-t/\tau}\,.
\label{eq.exponential}
\end{equation}
We calculate the time constants $\tau$ for the 20 subjects, and find
an average $\tau = 20 $ min with $\sigma=5$. 
Using the Kolmogorov-Smirnov test, we find that 
we cannot reject the null hypothesis that $p(t)$ of the sleep state of 
each subject of our 39 data sets is drawn from the same distribution
(Fig. ~\ref{f.pdf}b). We further find that $P(t)$ of the sleep state for 
the pooled data is consistent with an exponential distribution with a 
characteristic 
time $\tau = 22 \pm 1$ min (Fig.~\ref{f.data}b).


In order to verify that $P(t)$ of sleep state 
is better described by an exponential 
functional form rather than by a stretched exponential 
functional form, we fit these curves to the $P(t)$ from pooled 
data. Using Levenberg-Marquardt method, we find that the stretched 
exponential form lead to worse fit. 
The $\chi^2$ errors of exponential fit 
and stretched exponential fit are $8 \times 10^{-5}$ and $2.7 \times 10^{-2}$,
respectively.
We also check the results by plotting
$\log(|\log P(t)|)$ versus $\log t$
(\inst{1})
and find that the data are 
clearly more curved than when we plot $\log P(t)$ versus $t$,
indicating that an exponential form provides the best description of the data.

\begin{figure}
\begin{center}
\begin{minipage}[b]{0.4\linewidth}
 \epsfig{file=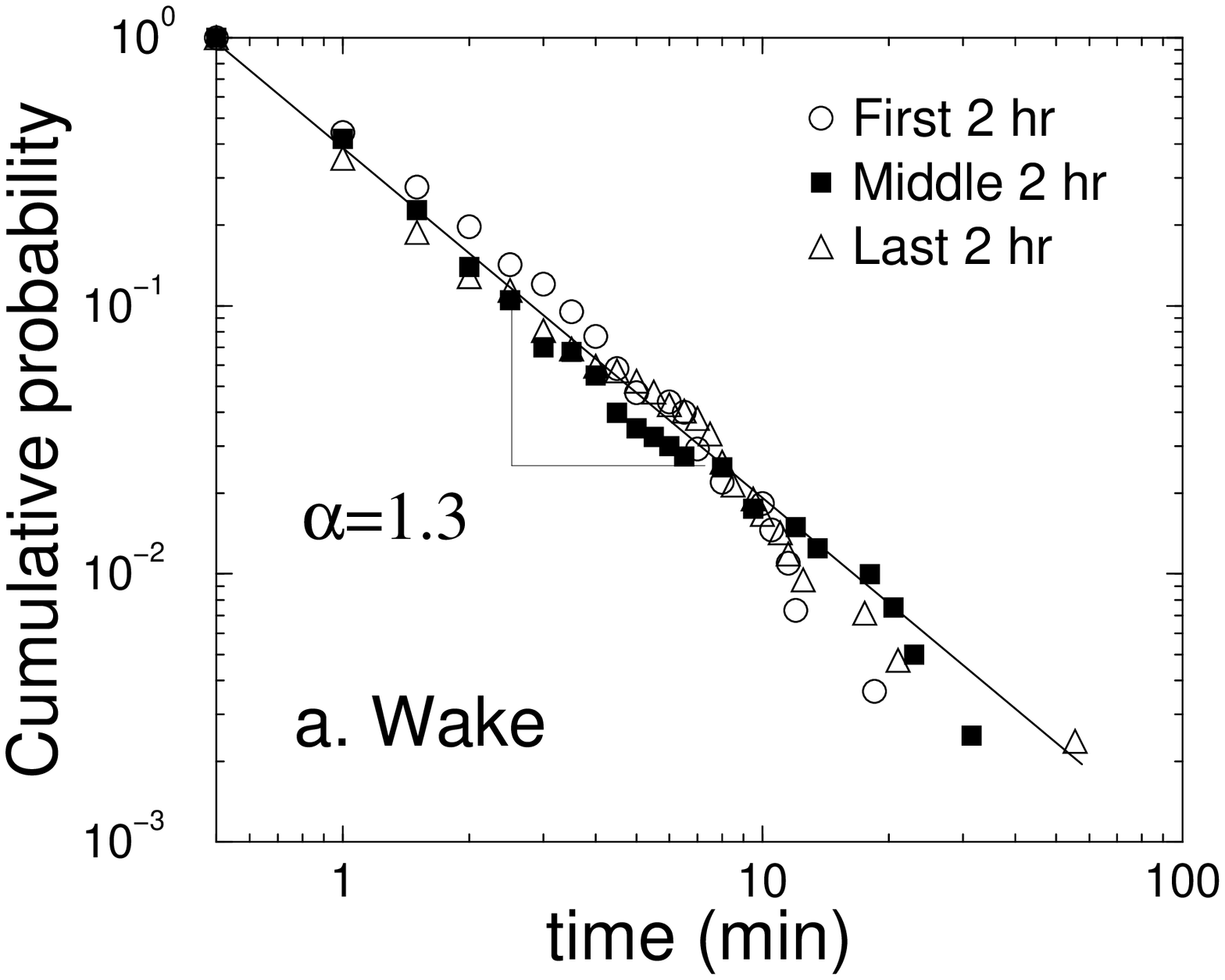, width=\linewidth}
\end{minipage} 
\begin{minipage}[b]{0.4\linewidth}
 \epsfig{file=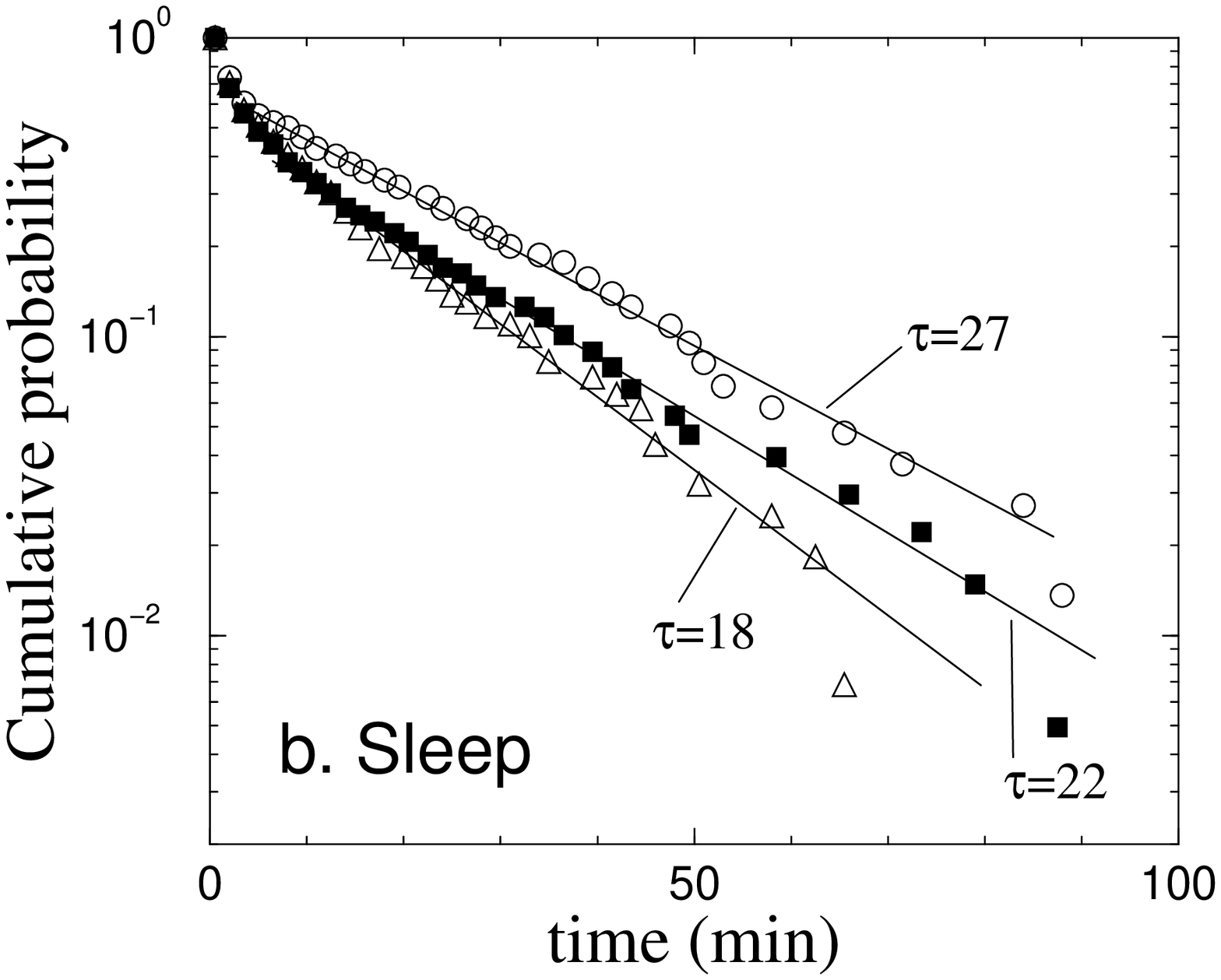, width=\linewidth}
\end{minipage}
\end{center}
\vspace{-0.4cm}
\caption{$P(t)$ of sleep and wake states in the first
two hours, middle two hours and last two hours of sleep. (a)
$P(t)$ of wake states;
the power-law exponent $\alpha$ does not change in a measurable way. 
(b) $P(t)$ of sleep states; 
the characteristic time $\tau$ decreases in the course of the night.
}
\vspace{-0.2cm}
\label{f.data.2hrs}
\end{figure}

Sleep is not a ``homogeneous process'' throughout the course of the night
\cite{Born-J-1999a,Carskadon-M-2000a},
so we ask if there is any change of $\alpha$ and $\tau$
during the night. We study sleep 
and wake durations for the first two hours, middle two hours, and the last 
two hours of nocturnal sleep using the pooled data from all 39 records 
(Fig.~\ref{f.data.2hrs}). Our results suggest that
$\alpha$ does not change for these three portions of the night, while 
$\tau$ decreases 
from $27 \pm 1$ min in the first two hours to $22 \pm 1$ min in the middle 
two hours, and then to $18 \pm 1$ min in the last two hours. The decrease 
in $\tau$ implies that 
the number of wake periods increases as the night proceeds, and we indeed find 
that the average number of wake periods for the last two hours is $1.4$ times 
larger than for the first two hours. 


\begin{figure}
\begin{center}
\begin{minipage}[b]{0.4\linewidth}
 \epsfig{file=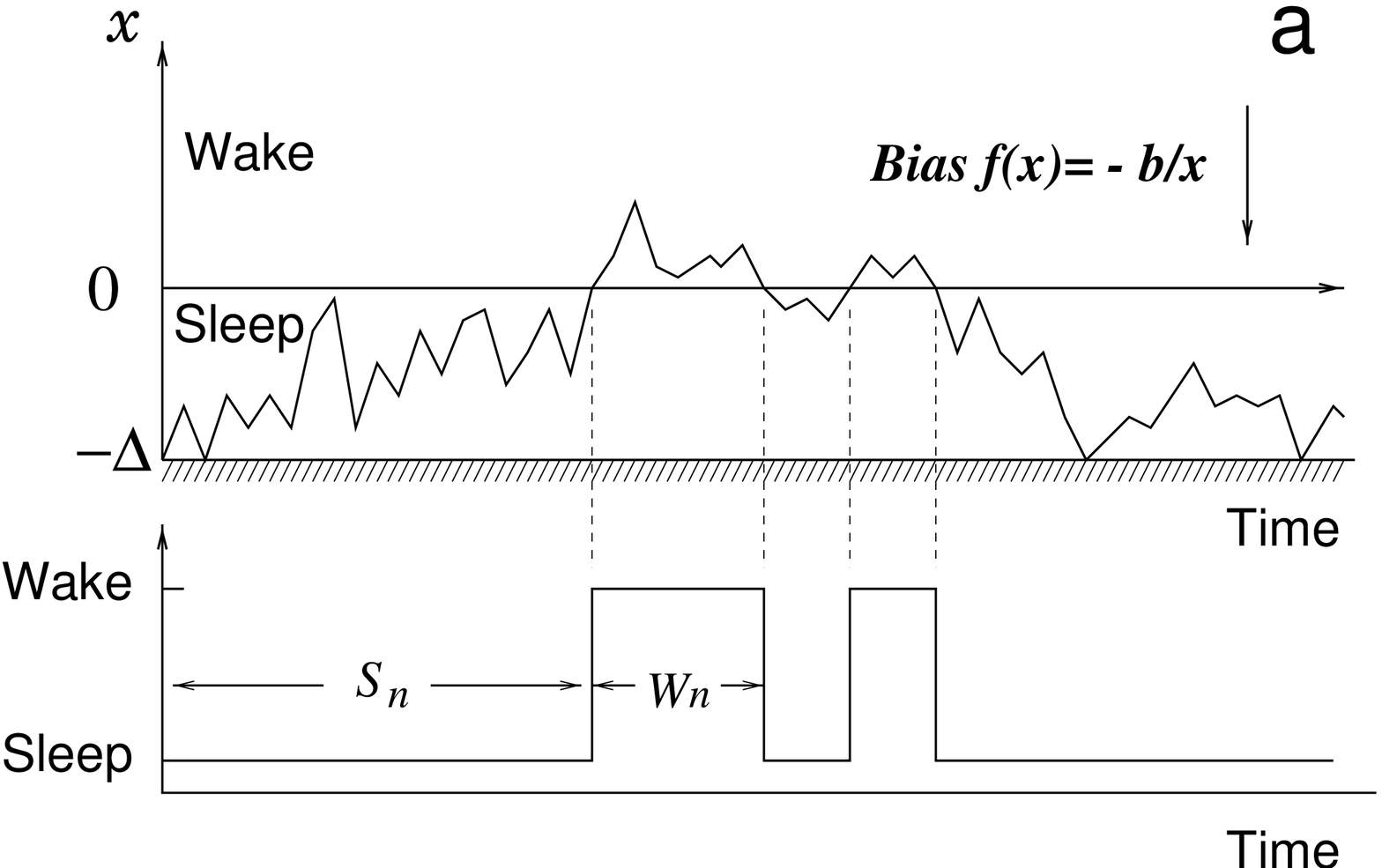, width=\linewidth}
\end{minipage} 
\begin{minipage}[b]{0.4\linewidth}
 \epsfig{file=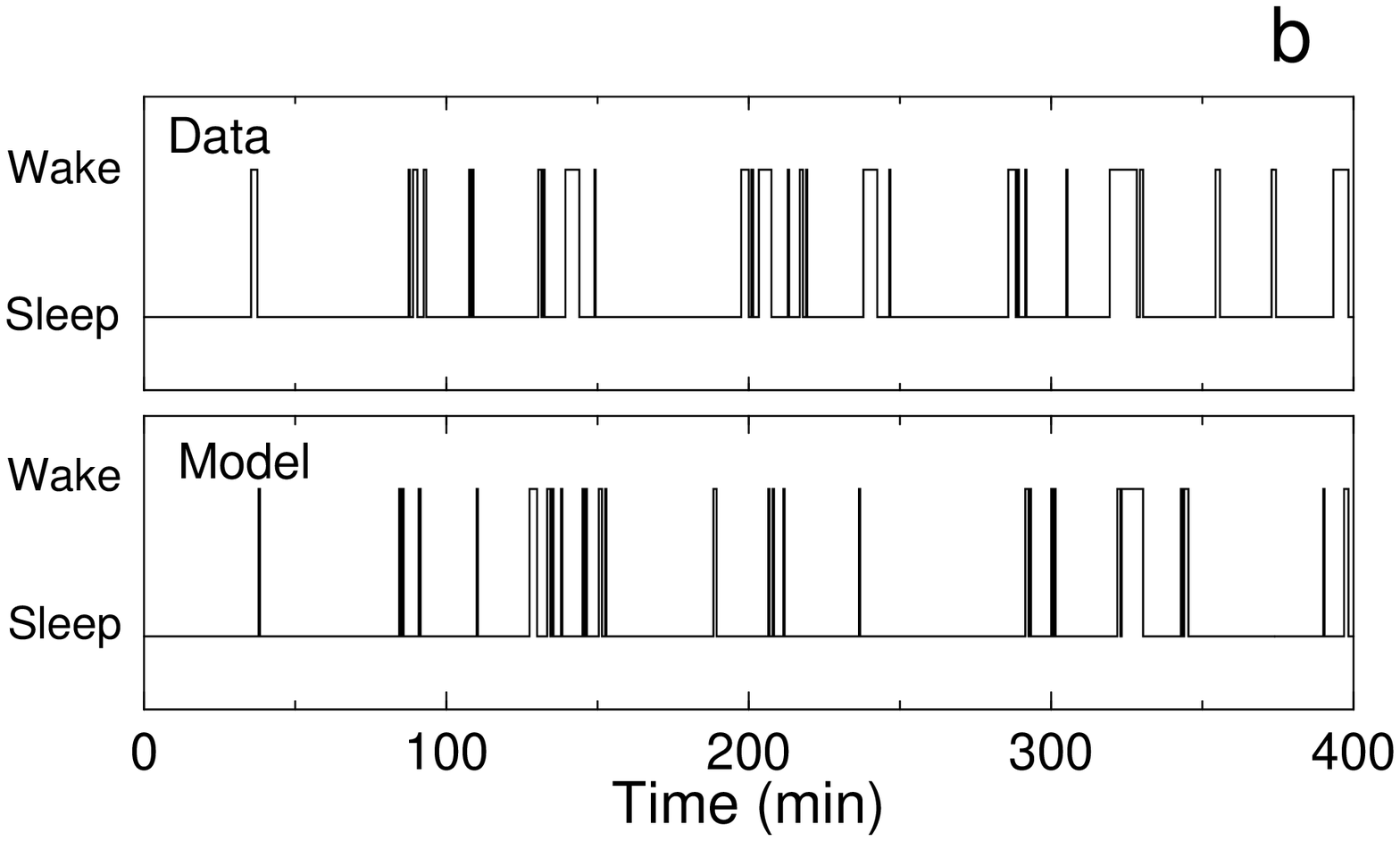, width=\linewidth}
\end{minipage}
\end{center}
\vspace{-0.4cm}
\caption{ (a) The upper panel illustrates the dynamics of the model. The
state $x(t)$ of the sleep-wake system evolves as a random walk with the
convention that $x>0$ corresponds to wake state and $-\Delta\leq x\leq 0$
corresponds to the sleep state.  In the wake state there is a ``restoring
force,'' $f(x)=-b/x$, ``pulling'' the system towards the sleep state.  The
lower panel illustrates sleep-wake transitions from the model, where 
$S_n$ and $W_n$ indicate the durations of the $n$-th sleep and
wake states, respectively. 
(b) Comparison of typical data and of a typical output of the model. The visual
similarity between the two records is confirmed by quantitative analysis
(Fig.~\protect\ref{f.model_results}).  }
\label{f.model}
\vspace{-0.2cm}
\end{figure}

We next investigate mechanisms that may be able to generate the different
behavior observed for sleep and wake.
Although several quantitative models, such as the two-process model 
\cite{Borbely-A-1999a} and the thermoregulatory model \cite{Nakao-M-1999a},
have been developed to describe human sleep regulation,
detailed modeling of frequent short awakening during nocturnal
sleep has not been addressed \cite{Dijk-D-1999a}.
To model the sleep-wake transitions, we make three assumptions
(Fig.~\ref{f.model}):

{\bf Assumption 1} defines the key variable $x(t)$
for sleep-wake dynamics. Although we consider a two-state
system, the brain as a neural system is unlikely to have only two discrete
states.  Hence, we assume that both wake and sleep ``macro'' states comprise
large number of ``microstates'' which we map onto a continuous
variable $x(t)$ defined in such a way that positive values correspond to
the wake state while negative values correspond to the sleep state.  We
further assume that there is a finite region $-\Delta \leq x \leq 0$ for the
sleep state.

{\bf Assumption 2} concerns the dynamics of the variable
$x(t)$.  Recent studies
\cite{McGinty-D-2000a,Gallopin-T-2000a}
suggest that a small population of sleep-active neurons in a localized
region of the brain distributes inhibitory inputs to wake-promoting neuronal
populations, which in turn interact through a feedback on the sleep-active
neurons. Because of these complex interactions, the global state of the 
system may present a ``noisy'' behavior.  Accordingly, we assume
that $x(t)$ evolves by a random-walk type of dynamics due to the competition
between the sleep-active and wake-promoting neurons.

{\bf Assumption 3} concerns a bias towards sleep.
We assume that if $x(t)$
moves into the wake state, then there will be a ``restoring force'' pulling
it towards the sleep state. This assumption corresponds to the common
experience that in wake periods during nocturnal sleep, one usually has
a strong tendency to quickly fall asleep again. Moreover, the longer one
stays awake, the more difficult it may be to fall back asleep, so we
assume that the restoring force becomes weaker as one moves away
from the transition point $x=0$.  We model these observations by assuming that the
random walker moves in a logarithmic 
potential $V(x)=b\ln x$, yielding a force $f(x)\equiv -dV(x)/dx=-b/x$, where the
bias $b$ quantifies the strength of the force.

Assumptions 1-3 can be written compactly as:
\begin{eqnarray}
\delta x(t)
\equiv x(t+1) - x(t)=
 \begin{cases}
  \epsilon(t), &\text{if}\; -\Delta \leq x(t) \leq 0 \qquad \text{(sleep)}, \\
 - \frac{b}{x} + \epsilon (t), &\text{if}\;\;\; x(t) > 0 \ \ \qquad \qquad
   \text{(wake)},
 \end{cases}
\label{eq.model}
\end{eqnarray}
where $\epsilon(t)$ is an uncorrelated Gaussian-distributed random
variable with zero mean and unit standard deviation.
In our model, the bias $b$ and the threshold $\Delta$ may change 
during the course of the night due to physiological variations
such as circadian cycle ~\cite{Borbely-A-1999a,Dijk-D-1999a}.


In our model, the distribution of durations of the wake state is
identical to the distribution of return times of a random 
walk in a logarithmic potential. For large times, this distribution 
is of a power law form
\cite{Zapperi-S-1998a,Havlin-S-1985a,ben-Avraham-D-2000a,Bray-A-2000a}.
Hence, for large times, the cumulative distribution of return times is 
also a power law, Eq.~(\ref{eq.power-law}), and the exponent is predicted
to be
\begin{equation}
 \label{eq.alpha}
 \alpha = \frac{1}{2} + b \,.
\end{equation}
From Eq.~(\ref{eq.alpha}) it follows that the cumulative distribution of
return times for a
random walk without bias ($b=0$) decreases as a power law with an exponent
$\alpha=1/2$. Note that introducing a
restoring force of the form $f(x)= - b/{x^\gamma}$ 
with $\gamma \neq 1$, yields stretched exponential distributions 
\cite{Bray-A-2000a}, so $\gamma = 1$ is
the only case yielding a power-law distribution.

\begin{figure}

\begin{center}
\begin{minipage}[b]{0.4\linewidth}
 \epsfig{file=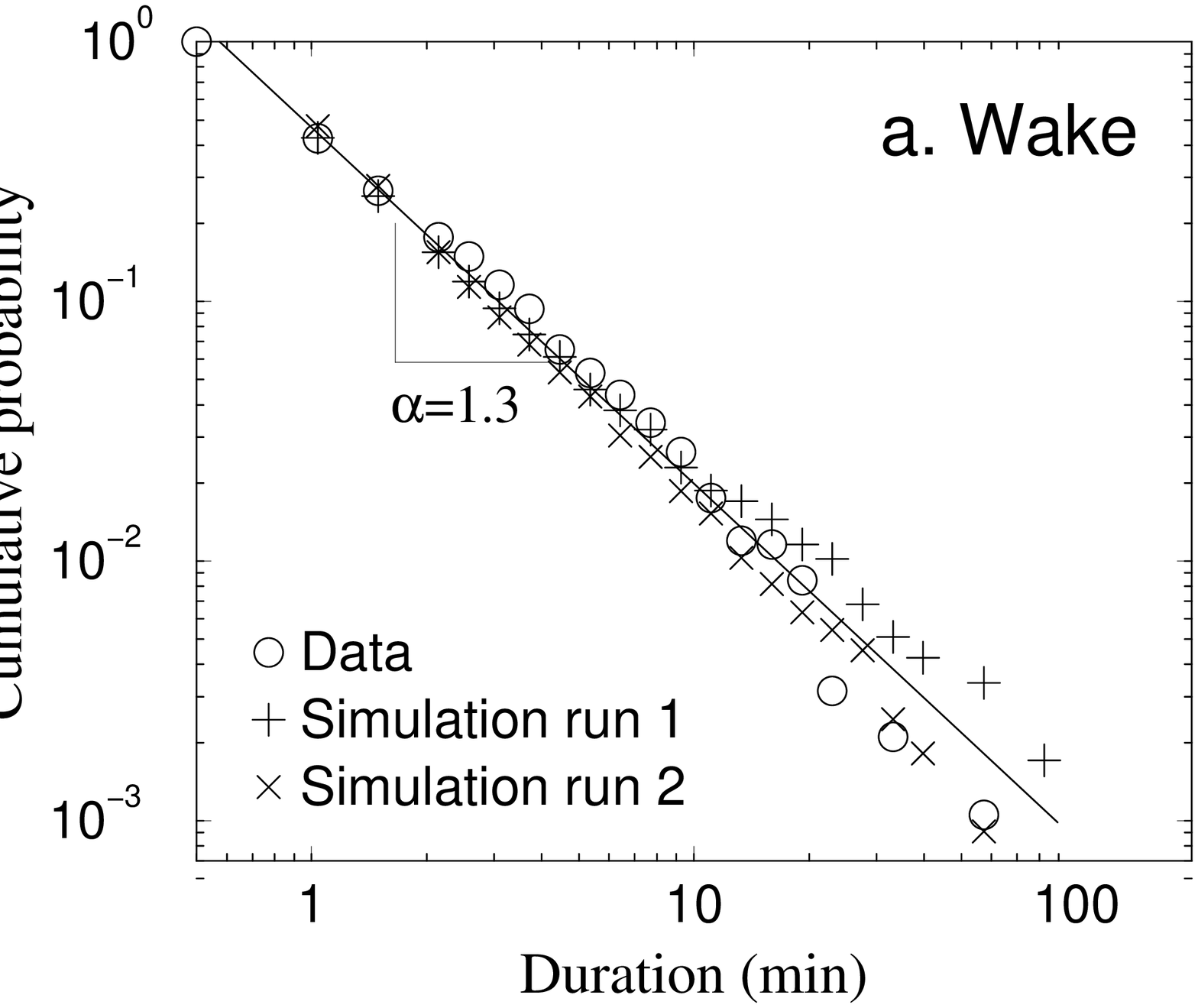, width=\linewidth}
\end{minipage} \hspace{0.5cm}
\begin{minipage}[b]{0.4\linewidth}
 \epsfig{file=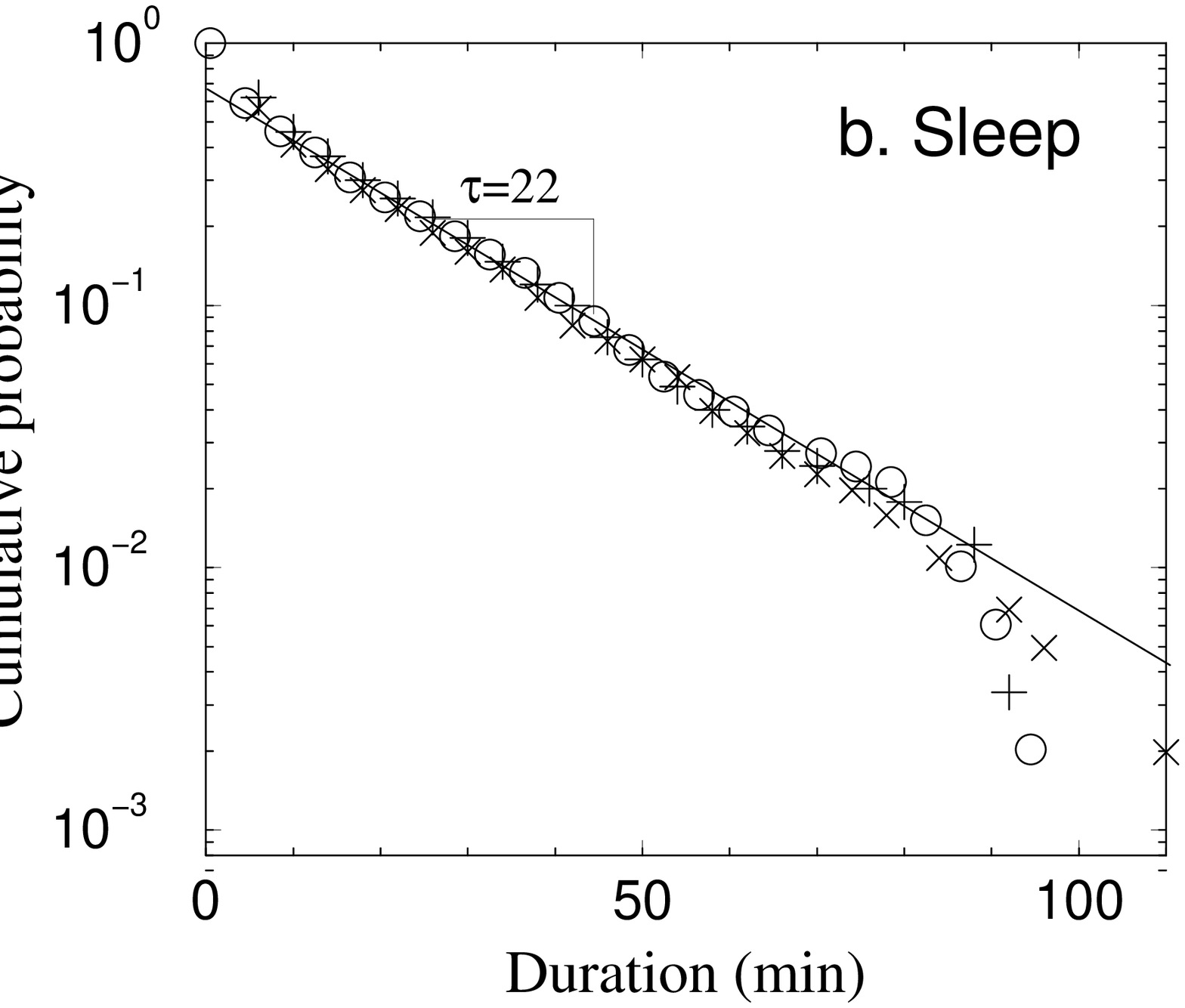, width=\linewidth}
\end{minipage}
\end{center}
\vspace{-0.4cm}
\caption{
Comparison of $P(t)$ for data and model (two runs with
same parameters). (a) $P(t)$ of the wake state. (b) $P(t)$ of the 
sleep state.
Note that the choice of $\Delta$ depends on the choice of the
time unit of the step in the model. We choose the time unit to be 30 seconds,
which corresponds to the time resolution of the data. To avoid big jumps in
$x(t)$ due to the singularity of the force when $x(t)$ approaches $x=0$,
we introduce a small constant $\lambda$ in the definition of the
restoring force $f(x)=-b/(x+\lambda)$.  
We find that the value of $\lambda$ does not change $\alpha$ or $\tau$.
}
\vspace{-0.2cm}
\label{f.model_results}
\end{figure}

Similarly, the distribution of durations of the sleep state is identical to
the distribution of return times of a random walk in a space with
a reflecting boundary.  Hence $P(t)$ has an
exponential distribution, Eq.~(\ref{eq.exponential}), in the large time 
region, with the
characteristic time $\tau$ predicted to be
\begin{equation}
 \label{eq.tau}
 \tau \sim \Delta^2 \,.
\end{equation}
Equations (\ref{eq.alpha}) and (\ref{eq.tau}) indicate that the values
of $\alpha$ and $\tau$ in the data can be reproduced in our model by
``tuning'' the threshold $\Delta$ and the bias $b$
(Fig.~\ref{f.model_results}). The decrease of the
characteristic duration of the sleep state as the night proceeds is consistent 
with the possibility that $\Delta$ decreases.
Our calculations suggest that $\Delta$
decreases from $7.9 \pm 0.2$ in the first hours of sleep, to $6.6 \pm 0.2$
in the middle hours, and then to $5.5 \pm 0.2$
for the final hours of sleep. Accordingly, the number of wake periods of 
the model increases by
a factor of 1.3 from the first two hours to the last two hours,
consistent with the data.
However, the apparent consistency of the power-law exponent for
the wake state suggests that the bias $b$ may remain
approximately constant during the night. Our best estimate is
$b=0.8 \pm 0.1$. 


To further test the validity of our assumptions, we examine the correlation
between the durations of consecutive states.  
Consider the sequence of sleep and wake durations \{~$S_1
~W_1 ~S_2 ~W_2 .... S_n ~W_n$~\}, where $S_n$ indicates the duration of
$n$-th sleep period and $W_n$ indicates the duration of $n$-th wake period
(Fig.~\ref{f.model}a).
Our model predicts that there are no autocorrelations in the series
$S_n$ and $W_n$, as well as no cross-correlations between series $S_n$
and $W_n$, the reason being that the uncorrelated random walk carries no
information about previous steps.  The experimental data confirms these 
predictions,
within statistical uncertainties.


Our findings of a power-law distribution for wake periods and 
an exponential distribution for sleep periods are intriguing because 
the same sleep-control mechanisms give rise to two completely different 
types of 
dynamics---one without a characteristic scale and the other with.
Our model suggests that the difference in the dynamics of the sleep 
and wake states
(e.g. power law versus exponential) arises from the distinct number of 
microstates that can be explored by the sleep-wake system for these two states.
During the sleep state, the system is confined in the region 
$-\Delta \leq x \leq 0 $. The parameter $\Delta$ imposes a scale which 
causes an exponential distribution of durations.
In contrast, for the wake state the system can 
explore the entire half-plane $x>0$. 
The lack of constraints leads to a scale-free power-law distribution of
durations. In addition, the $1/x$ restoring force in the wake state 
does not change the functional form of the distribution,
but its magnitude determines the power-law exponent of the
distribution using Eq. (\ref{eq.alpha}).

Although in our model the sleep-wake system can explore 
the entire half-plan $x>0$ during wake periods, the
``real'' biological system is unlikely to generate very large
value (i.e., extreme long wake durations). There must 
be a constraint or boundary in the wake state at a
certain value of $x$. If such a constraint or boundary exists, we 
will find a cut-off with exponential tail in the distribution
of durations of the wake state. More data are needed to test this 
hypothesis.

Our additional finding of a stable power-law behavior for 
wake periods for all portions of the night implies that the mechanism 
generating the 
restoring force in the wake state is not affected in a measurable way 
by the mechanism
controlling the changes in the durations of the sleep state. We hypothesize 
that even though the power-law behavior does not change in the course 
of the night for healthy individuals, it may change under pharmacological 
influences or under different conditions, such as stress or depression. 
Thus, our results may also be
useful for testing these effects on the statistical 
properties of the wake state and the sleep state.
\begin{center}
***
\end{center}
We thank the NIH/National Center for Research Resource 
(P41 RR 13622) for support and A. L. Goldberger and C.-K. 
Peng for discussions.


\end{document}